\documentclass[pdflatex,sn-mathphys-num]{sn-jnl}

\usepackage{graphicx}%
\usepackage{multirow}%
\usepackage{amsmath,amssymb,amsfonts}%
\usepackage{amsthm}%
\usepackage{mathrsfs}%
\usepackage[title]{appendix}%
\usepackage{xcolor}%
\usepackage{textcomp}%
\usepackage{manyfoot}%
\usepackage{booktabs}%
\usepackage{algorithm}%
\usepackage{algorithmicx}%
\usepackage{algpseudocode}%
\usepackage{listings}%
\usepackage{natbib}

\theoremstyle{thmstyleone}%
\theoremstyle{thmstyletwo}%

\theoremstyle{thmstylethree}%

\begin{document}

\title[Article Title]{DiffRenderGAN: Addressing Training Data Scarcity in Deep Segmentation Networks for Quantitative Nanomaterial Analysis through Differentiable Rendering and Generative Modelling}


\author[1,5]{\fnm{Dennis} \sur{Possart}}
\author[2, 9]{\fnm{Leonid} \sur{Mill}}
\author[3]{\fnm{Florian} \sur{Vollnhals}}
\author[4]{\fnm{Tor} \sur{Hildebrand}}

\author[4]{\fnm{Peter} \sur{Suter}}
\author[2]{\fnm{Mathis} \sur{Hoffmann}}
\author[1]{\fnm{Jonas} \sur{Utz}}
\author[5]{\fnm{Daniel} \sur{Augsburger}}
\author[2]{\fnm{Mareike} \sur{Thies}}
\author[2]{\fnm{Mingxuan} \sur{Wu}}
\author[2]{\fnm{Fabian} \sur{Wagner}}
\author[3,5,7]{\fnm{George} \sur{Sarau}}
\author[3,5,6,7]{\fnm{Silke}\sur{Christiansen}}
\author[1, 8]{\fnm{Katharina} \sur{Breininger}}

\affil[1]{\orgdiv{Department Artificial Intelligence in Biomedical Engineering}, \orgname{Friedrich-Alexander-University Erlangen-Nürnberg}, \postcode{91052}, \state{Erlangen}, \country{Germany}}

\affil[2]{\orgdiv{Pattern Recognition Lab}, \orgname{Friedrich-Alexander-University Erlangen-Nürnberg}, \postcode{91058}, \state{Erlangen}, \country{Germany}}

\affil[3]{\orgdiv{Institute for Nanotechnology and Correlative Microscopy}, \postcode{91301}, \state{Forchheim}, \country{Germany}}

\affil[4]{\orgdiv{Lucid Concepts AG}, \postcode{8005}, \state{Zurich}, \country{Switzerland}}

\affil[5]{\orgdiv{Correlative Microscopy and Materials Data}, \orgname{Fraunhofer Institute for Ceramic Technologies and Systems}, \postcode{91301}, \state{Forchheim}, \country{Germany}}

\affil[6]{\orgdiv{Institute of Experimental Physics}, \orgname{Freie Universität Berlin}, \postcode{91301}, \state{Berlin}, \country{Germany}}

\affil[7]{\orgdiv{Emeritus-Gruppe Leuchs}, \orgname{Max Planck Institute for the Science of Light}, \postcode{91058}, \state{Erlangen}, \country{Germany}}

\affil[8]{\orgdiv{Center for AI and Data Science}, \orgname{Julius-Maximilians-University Würzburg}, \postcode{97074}, \state{Würzburg}, \country{Germany}}

\affil[9]{\orgdiv{MIRA Vision Microscopy GmbH},  \postcode{73037}, \state{Göggingen},\country{Germany}}


\abstract{
Nanomaterials exhibit distinctive properties governed by parameters such as size, shape, and surface characteristics, which critically influence their applications and interactions across technological, biological, and environmental contexts. Accurate quantification and understanding of these materials are essential for advancing research and innovation. In this regard, deep learning segmentation networks have emerged as powerful tools that enable automated insights and replace subjective methods with precise quantitative analysis. However, their efficacy depends on representative annotated datasets, which are challenging to obtain due to the costly imaging of nanoparticles and the labor-intensive nature of manual annotations. To overcome these limitations, we introduce DiffRenderGAN, a novel generative model designed to produce annotated synthetic data. By integrating a differentiable renderer into a Generative Adversarial Network (GAN) framework, DiffRenderGAN optimizes textural rendering parameters to generate realistic, annotated nanoparticle images from non-annotated real microscopy images. This approach reduces the need for manual intervention and enhances segmentation performance compared to existing synthetic data methods by generating diverse and realistic data. Tested on multiple ion and electron microscopy cases, including titanium dioxide (TiO\(_2\)), silicon dioxide (SiO\(_2\)), and silver nanowires (AgNW), DiffRenderGAN bridges the gap between synthetic and real data, advancing the quantification and understanding of complex nanomaterial systems.
}

\maketitle

\section{Main}\label{main}
Nanomaterials are ubiquitous and exhibit unique properties that are often dictated by their size, shape, and surface characteristics. These attributes influence not only their performance in technological applications but also their interactions within biological and environmental systems. A precise understanding of these parameters is therefore critical across fields, whether the goal is to optimize material properties for advanced technologies or to assess potential risks in environmental and health contexts. For example, titanium dioxide (TiO\(_2\)) and silicon dioxide (SiO\(_2\)) nanoparticles are used in a wide range of applications, from nanomedicine \cite{wu2020tio2}\cite{huang2022silica} to photocatalysis \cite{gupta2011review} and wastewater treatment \cite{nayl2022recent}. Furthermore, silver nanowires (AgNWs) are promising candidates for indium-free transparent electrodes \cite{goebelt2015encapsulation}\cite{han2015fully}. 

To effectively analyze nanomaterials, automated methods are necessary, particularly when dealing with complex particle agglomerates and large numbers of particles. Deep learning segmentation networks have emerged as powerful tools in this regard, transforming quantitative analysis in microscopic imaging from traditional subjective methods to precise and automated approaches \cite{ronneberger2015unetconvolutionalnetworksbiomedical}. For example, these networks now offer unprecedented insight into pathological findings \cite{van2021deep}\cite{marc_aubreville_2022_6362337} and material production processes \cite{fu2022deep}\cite{shammaa2010segmentation}.

However, their ability to generalize to novel, unseen data critically depends on the availability of representative training datasets \cite{zhang2021understanding}, as these datasets determine the data distribution from which diverse and class defining features are derived \cite{bishop2006pattern}. If the training data distribution insufficiently represents the problem at hand, models will perform unsatisfactorily \cite{bishop2006pattern}\cite{theodoridis2006pattern}. In microscopic imaging, several challenges hinder the acquisition of comprehensive datasets, including high equipment costs, reliance on highly specialized personnel, and the labor-intensive nature of manual image annotation. 

To address these challenges, researchers have increasingly turned to data synthesis methods. Generative adversarial networks (GANs) have shown significant potential in generating synthetic annotated data in an unsupervised manner, effectively capturing the essence of real data \cite{goodfellow2014generative}\cite{frid2018synthetic}\cite{jordon2018pate}\cite{guan2019breast}. For example, Rühle et al. (2021) \cite{ruhle2021workflow} successfully utilized WassersteinGANs \cite{arjovsky2017wassersteingan} and CycleGANs \cite{CycleGAN2017} to synthesize annotated Scanning Electron Microscopy (SEM) images for the identification and segmentation of TiO\(_2\) nanoparticles. Other approaches have explored the incorporation of prior knowledge into the data synthesis process \cite{maier2019learning}, such as expert-guided image rendering \cite{wood2021faketillmakeit}\cite{mill2021synthetic}. Mill et al. (2021) \cite{mill2021synthetic} demonstrated this technique by simulating Helium-Ion Microscopy (HIM) images of SiO\(_2\) and TiO\(_2\) nanoparticles to train expressive segmentation networks. 

Although synthetic data was effectively used in the studies of Rühle et al. (2021) and Mill et al. (2021), evaluation results showed that segmentation models trained on synthetic data generally underperformed in most metrics compared to those trained on real data, indicating a domain gap in synthetic data \cite{mill2021synthetic}\cite{ruhle2021workflow}. For the GAN-based method of Rühle et al. (2021), reduced segmentation performance could be attributed to factors such as visual artifacts, training instability, and inaccuracies in the synthetic labels. In contrast, Mill et al.'s (2021) rendering approach may have exhibited lower segmentation performance due to the omission of class-important features that exceed the identification and rendering capabilities of domain experts.

\begin{figure}[t]
\centering
\includegraphics[width=0.9\textwidth]{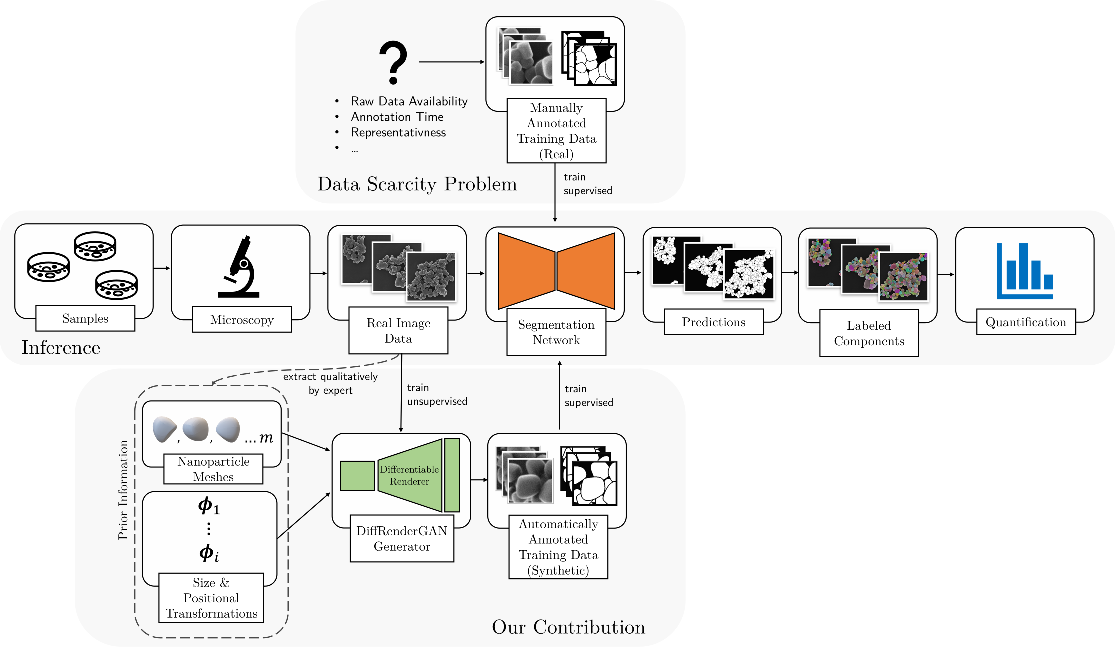}
\caption{\textbf{Addressing Training Data Scarcity in Deep Segmentation Networks for Quantitative Nanomaterial Analysis through Synthetic Data Generation.} Our contribution aims to address three primary objectives: (1) to present an image synthesis method applicable across various microscopy modalities for the analysis of materials with diverse morphologies, (2) to minimize the need for expert intervention, and (3) to reduce or eliminate the representativeness gap between synthetic and real data, as observed in previous studies, enabling more efficient training of deep segmentation networks for improved analysis of complex nanomaterial systems. It is important to note that our goal is not to generate a physically accurate simulation of materials but rather to conduct a simulation that produces images capturing the characteristics necessary for training a generalizing segmentation network.}
\label{fig:main}
\end{figure}

Recent advances in differentiable rendering offer new potential by enabling the automatic optimization of reality-replicating 3D models using gradient descent methods \cite{Jakob2020DrJit}. This minimizes reliance on manual expertise while enhancing the realism of synthetic images. Building on this potential and the unsupervised training capabilities of GANs, we combine both techniques and introduce DiffRenderGAN, a novel generative model that integrates a differentiable renderer within a GAN framework. Using nanoparticle 3D models, such as meshes, and a transformation matrix containing positional and scaling information to arrange these meshes realistically, DiffRenderGAN learns distributions of textural rendering parameters that simulate materials from a given real nanoparticle dataset. This parametric representation enables the generation of synthetic, annotated images that closely mirror real measured data. These images can then be used to train segmentation networks effectively, facilitating the identification and quantification of nanoparticles in measured microscopy images.

In Figure \ref{fig:main}, we summarize the contributions of this work. This paper presents DiffRenderGAN and demonstrates its application across various microscopy datasets, including those from Mill et al. (2021) \cite{mill2021synthetic} (SiO\(_2\), TiO\(_2\) in HIM) and Rühle et al. (2021) \cite{ruhle2021workflow} (TiO\(_2\) in SEM), as well as a silver nanowire (AgNW) dataset using multi-beam SEM. We evaluate DiffRenderGAN by comparing the synthetic data it generates with other methods, training segmentation models only on synthetic data, and testing them on real microscopy images. For the AgNW dataset, where ground truth annotations are unavailable, we assess DiffRenderGAN qualitatively. 

Our results demonstrate that DiffRenderGAN effectively optimizes parameters for realistic image generation, reducing manual effort to selecting target meshes and training parameters. Our method meets or exceeds the performance of existing methods, achieving higher scores across key segmentation metrics. These results highlight the potential of DiffRenderGAN as a powerful tool for generating synthetic multimodal microscopy data, reducing the domain gap in synthetic images, and advancing the analysis and understanding of complex nanomaterial systems.

\section{Leveraging Differentiable Rendering for Enhanced Generative Modeling}
Our image synthesis method integrates the principles of image rendering and GANs. Image rendering involves transforming a virtual 3D scene into a realistic 2D digital image from a specified perspective \cite{pharr2023physically}. The virtual 3D scene is defined by parameters such as meshes (e.g., 3D nanoparticle models) and textures attached to them, referred to in this work as Bidirectional Scattering Distribution Functions (BSDFs), which simulate material properties like diffuse or dielectric behavior. In addition, light sources are included to define observable emissions. Formally, we denote the rendering process as \( f_{\text{r}} \), which generates an image \( I_{\text{r}} \) from the virtual scene expressed by \(\Theta\):

\begin{equation}
I_{\text{r}} = f_{\text{r}}(\Theta).
\end{equation}

The interested reader is referred to Kajiya et al. (1986) \cite{kajiya1986rendering} for a detailed definition and description of the rendering function \( f_{\text{r}} \). Rendering has been used in computer vision to create synthetic datasets for training machine learning models by integrating expert knowledge into the design of the virtual scene \cite{wood2021faketillmakeit}\cite{mill2021synthetic}\cite{rozantsev2015rendering}. 

One key advantage of rendering-based synthetic data is that annotation masks can be automatically extracted using unique identifiers assigned to each mesh in the virtual scene. Expert-guided image rendering prevents visual artifacts and labeling inaccuracies that are common in CycleGAN applications \cite{yoo2020cyclegan}\cite{osakabe2021cyclegan}\cite{de2021residual}. However, the expert-driven process of creating synthetic images is time-consuming, and key features might be overlooked in complex reference images. Therefore, a data-driven approach may be more desirable.

Differentiable rendering makes this possible by enabling the calculation of 
\(\frac{\partial I_{\text{r}}}{\partial \Theta}\), allowing for iterative optimization of virtual scene parameters \cite{Loubet2019Reparameterizing}. Using methods such as Stochastic Gradient Descent (SGD) or Adam \cite{robbins1951stochastic}\cite{kingma2017adammethodstochasticoptimization}, parameters can be adjusted to minimize an objective function, such as the Mean-Squared Error (MSE) between a rendered image and a target image. Replicating real data using a differentiable renderer presents significant challenges, particularly when working with large datasets of nanoparticle images. Achieving a realistic representation of each observed nanoparticle in images necessitates the accurate reconstruction and positioning of meshes, a process that becomes increasingly complex as the number of particles in the dataset grows. To address this challenge, we employ Generative Adversarial Networks (GANs), which are capable of generating realistic and diverse data distributions rather than exact replicas.

\begin{figure}[t]
\centering
\includegraphics[width=0.75\textwidth]{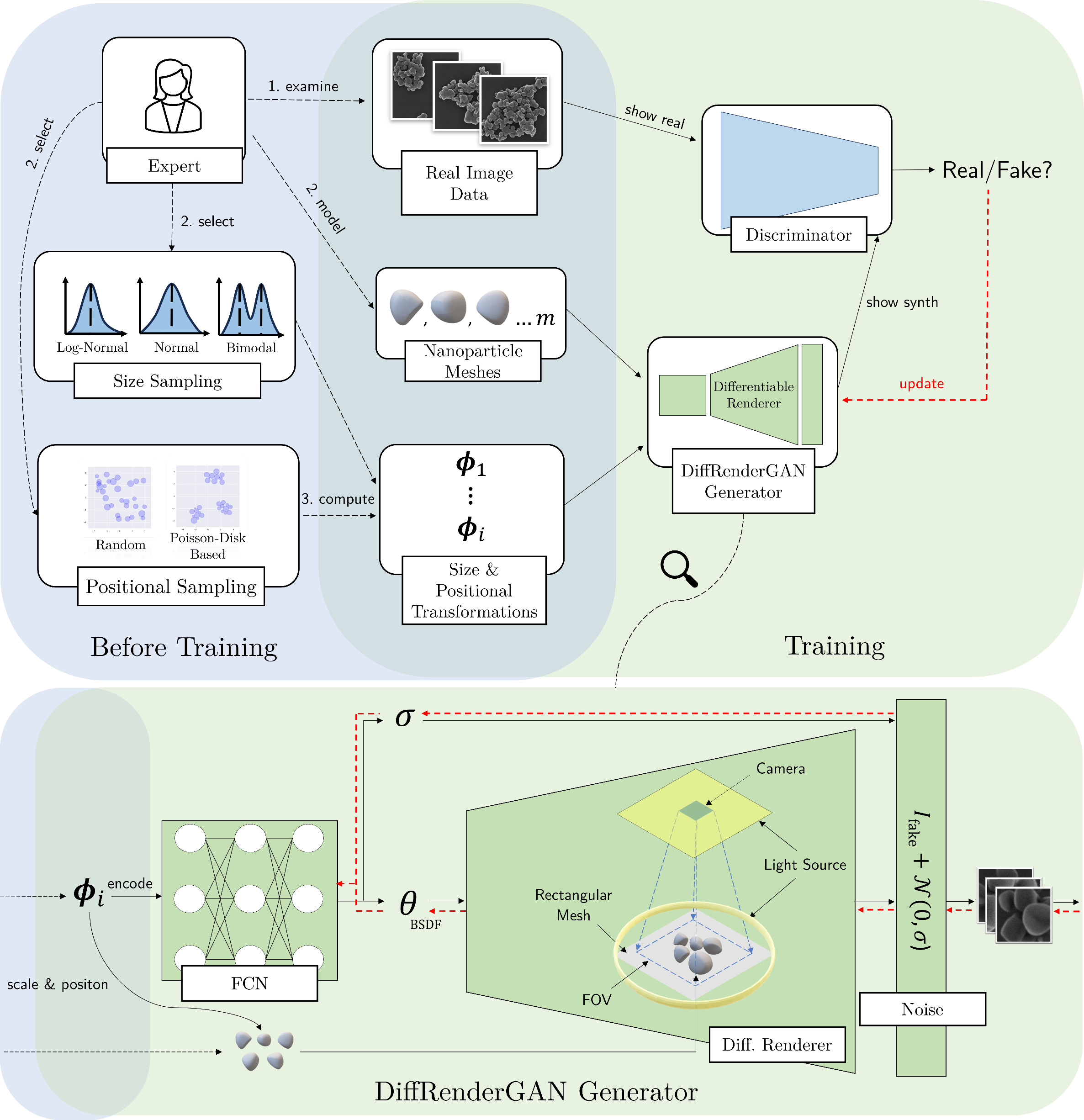}
\caption{\textbf{DiffRenderGAN Training Procedure.} Domain experts create target nanomaterial meshes to match the morphology of real particle systems. Scale and placement parameters are used to compute a transformation matrix for training. The meshes and transformation matrix serve as input to the DiffRenderGAN model. During image generation, a slice of the matrix is processed by a 5-layer Fully Connected Network (FCN) to predict BSDF parameters and noise scale. These parameters are passed to a differentiable renderer, which uses a virtual scene with scaled and positioned meshes to create the final synthetic nanomaterial image. A technical description of DiffRenderGAN's modules is provided in Section~\ref{m_model_design}. For visualization purposes, the virtual scene used by the differentiable renderer is shown in a simplified form. The actual structure can be found in the supplementary information of this paper, with a technical summary stated in Section~\ref{m_virtual_scene}.}
\label{fig:main2}
\end{figure}

GANs, introduced by Goodfellow et al. (2014) \cite{goodfellow2014generative}, consist of two neural networks: a generator \( G(z) \) that maps a random noise vector \( z \) from a distribution \( p_z \) into a synthetic image, and a discriminator \( D(x) \) that classifies images as real or fake, where \( x \) denotes a real sample from the distribution \( p_{\text{data}} \). The generator aims to produce images that are indistinguishable from real data, while the discriminator is tasked with effectively differentiating between real and synthetic images. The adversarial process is formulated as a min-max optimization problem:

\begin{equation}
\min_{G} \max_{D} \mathbb{E}_{x \sim p_{\text{data}}(x)} \left[ \log D(x) \right] + \mathbb{E}_{z \sim p_z(z)} \left[ \log(1 - D(G(z))) \right]
\label{vanilla_gan}
\end{equation}

By combining the unsupervised training capabilities of GANs with the controllability of differentiable rendering and automatic mask extraction, we developed DiffRenderGAN, which integrates a differentiable renderer into the GAN’s generator. This integration enables the generation of highly realistic synthetic images achieved without reconstruction. Simultaneously, the controlled rendering environment mitigates common visual artifacts, such as checkerboard patterns often observed in CycleGAN applications \cite{CycleGAN2017}, thereby ensuring higher-quality and more consistent outputs.

Optimizing all virtual scene parameters \(\Theta\) to visually simulate real nanoparticles, including their morphologies, is computationally demanding. To simplify this process, our generator focuses on optimizing textural parameters \(\theta_{\text{BSDF}}\) that mimic the material properties observed in SEM and HIM imaging. Assumptions regarding morphologies, size distribution, and placement of reference nanomaterials are provided by experts before training to guide DiffRenderGAN. This assumption-based strategy allows for a realistic arrangement of meshes without the need for direct optimization of their shapes and positions. We define the virtual scene parameter space \(\Theta\) as:

\begin{equation}
\Theta = \begin{pmatrix} \theta_{\text{BSDF}} \\ \theta_{\text{other}} \end{pmatrix},
\end{equation}

where \(\theta_{\text{BSDF}}\) includes all optimized BSDF parameters, while \(\theta_{\text{other}}\) encompasses non-optimizable BSDF parameters and all other scene parameters, including those related to geometry, position, and size.

Before training DiffRenderGAN (see Figure~\ref{fig:main2}), an expert-guided process is employed to model a collection of \( n \) particle meshes that reflect the shape properties of nanoparticles observed in real images (detailed in Section~\ref{m_mesh}). The sizes and positional arrangements of the meshes are selected from distributions such as normal, lognormal, or bimodal. Mesh placement can either be random or agglomerated, utilizing a Poisson Disk-based sampling algorithm for clustering \cite{bridson2007fast}. Subsequently, based on the selected placement and scale strategy, a transformation tensor 

\begin{equation}
\Phi = \{ \phi_i \mid \phi_i \in \mathbb{R}^{n \times 4}, \ i = 0, 1, \ldots, m-1 \}
\end{equation}

is computed, where each subtensor \(\phi_i\) contains spatial coordinates and a scaling factor for each of the \( n \) meshes. The tensor \(\Phi\) encodes \( m \) different nanoparticle constellations, defining the synthetic image sampling size with varying mesh arrangements used during training. We detail the computation of the transformation tensor in Section~\ref{m_trafo}.

\begin{figure}[t]
\centering
\includegraphics[width=1.\textwidth]{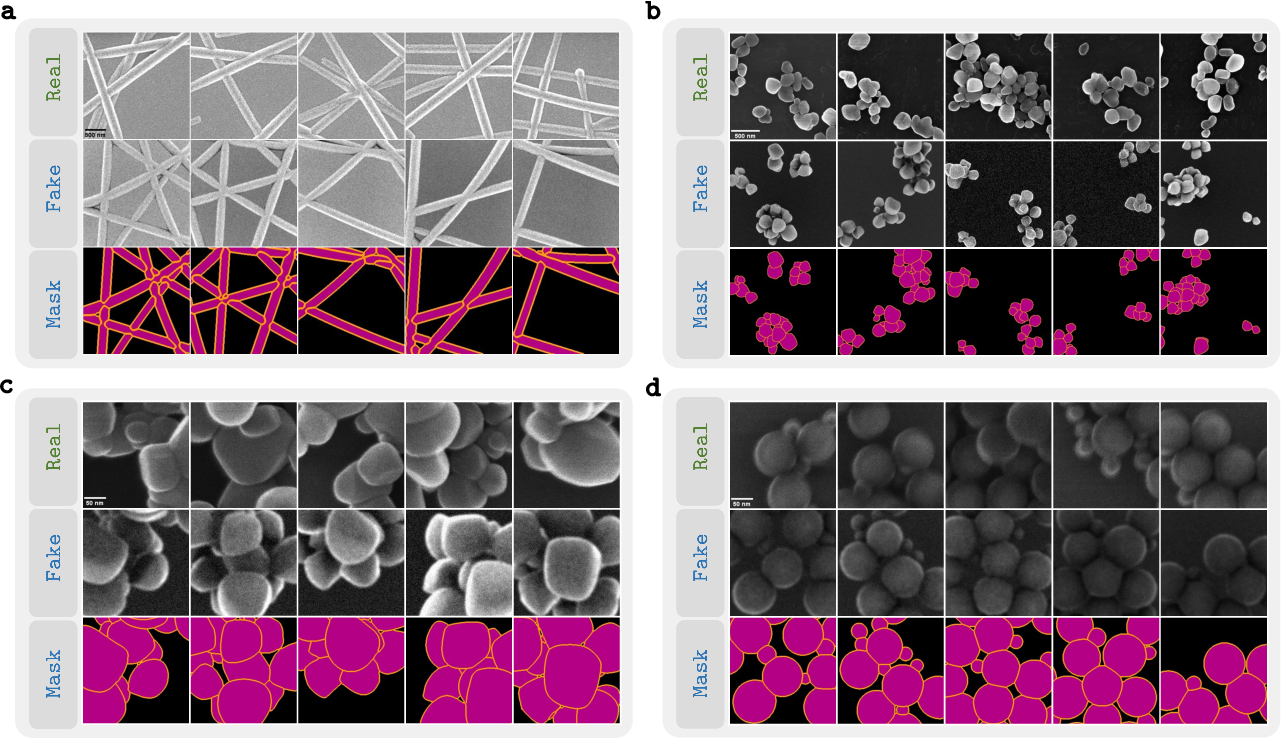}
\caption{\textbf{Comparison of Real and Synthetic Image Patches with Corresponding Segmentation Masks.} 
In each figure section, the top row shows real images used to train DiffRenderGAN, the middle row depicts synthetic images, and the bottom row shows the corresponding segmentation masks, highlighting material classes (purple) and boundaries (orange). These synthetic image-mask pairs serve as training data for multiclass segmentation networks as demonstrated in Section~\ref{Evaluation}. \textbf{a.} AgNW: trained using 10 bent cone meshes, choosing for transformation computation random placement in 2D and a lognormal size distribution.  
\textbf{b.} TiO$_2$ in SEM from Rühle et al. (2021) \cite{ruhle2021workflow}: trained using 40 cubically deformed meshes, choosing for transformation computation Poisson Disk-based placement in 3D and a lognormal size distribution.  
\textbf{c.} TiO$_2$ in HIM from Mill et al. (2021) \cite{mill2021synthetic}: trained using 15 cubically deformed meshes, choosing for transformation computation Poisson Disk-based placement in 3D and a lognormal size distribution.  
\textbf{d.} SiO$_2$ in HIM from Mill et al. (2021) \cite{mill2021synthetic}: trained using 20 sphere meshes, choosing for transformation computation Poisson Disk-based placement in 3D and a lognormal size distribution.}
\label{fig:synth}
\end{figure}

The architecture of DiffRenderGAN's generator is organized into three modules, as depicted in Figure~\ref{fig:main2}. The first module, a Fully Connected Network (FCN), denoted as \( f_{\text{fcn}} \), takes a uniformly randomly selected \(\phi_i \sim \mathcal{U}(\Phi)\), serving as a distinct mapping. This is analogous to the role of the randomly sampled noise vector \( z \) in vanilla GANs (as illustrated in Equation~\eqref{vanilla_gan}). The FCN regresses \(\theta_{\text{BSDF}}\) and a noise scale \(\sigma\), which is subsequently used in the generator's final module to introduce learnable Gaussian noise:

\begin{equation}
    (\theta_{\text{BSDF}}, \sigma) = f_{\text{fcn}}(\phi_i).
\end{equation}

For regularization and training stability, \(f_{\text{fcn}}\) produces parameter estimates in \([0,1]\). We then rescale and clip these values for both the BSDF parameters \(\theta_{\text{BSDF}}\) and the noise deviation \(\sigma\) so that they lie in their respective physically and render-environmental feasible ranges. The specific limits, including noise, stage, and particle mesh BSDF boundaries, are listed in the supplementary information. In the second module, a virtual scene is dynamically created where the collection of \( n \) expert-generated particle meshes is positioned and scaled according to \(\phi_i\). The latest BSDF values from \(\theta_{\text{BSDF}}\) are applied to both the nanoparticle meshes and a rectangular stage mesh located beneath them. The virtual scene is then passed to the differentiable renderer, \( f_{\text{r}} \), to generate a synthetic image. 

To simulate real-world imaging conditions, the third module, \( f_{\text{noise}} \), adds zero-centered Gaussian noise scaled by \(\sigma\) to the rendered image. This step aims to replicate the noise present in real images, making the synthetic output more realistic. The final synthetic image \( I_{\text{synth}} \) is evaluated by the discriminator for its realism, allowing for gradient computation via backpropagation to update weights in the generator's \( f_{\text{fcn}} \) \cite{rumelhart1986learning}. The generator's overall functionality is summarized as:

\begin{equation}
I_{\text{synth}} = G(\phi_i) = f_{\text{noise}}(f_{\text{r}}(f_{\text{fcn}}(\phi_i), \theta_{\text{other}}), \sigma).
\end{equation}

Within the adversarial training framework, DiffRenderGAN's generator and discriminator engage in the following adversarial process:

\begin{equation}
\min_{G} \max_{D} \mathbb{E}_{x \sim p_{\text{data}}(x)}[\log D(x)] + \mathbb{E}_{\phi_i \sim \mathcal{U}(\Phi)}[\log (1 - D(G(\phi_i)))].
\end{equation}

The available data from each image case introduced in Section~\ref{main} is split into approximately 80\% for training and 20\% for testing (later used in Section~\ref{Evaluation}). Details of the image data acquisition and a sample description are provided in Section~\ref{method_data}. Each image of the training dataset is cropped into overlapping patches of size \( 256 \times 256 \) pixels. DiffRenderGAN is then trained on image patches that contain at least three fully displayed particles while avoiding repetitive particle patches. 

At the same time, we demonstrate an effective use of image patches that do not contain particles but still provide valuable background information, such as artifacts, which do not necessitate additional annotation for particle segmentation tasks. We extract 200 of these patches for each dataset, which we later use to supplement our synthetic datasets.

The training process is monitored using the Fréchet Inception Distance (FID) score, a state-of-the-art metric that measures the feature distance between the generated synthetic images and real images \cite{parmar2021cleanfid}. To determine the best epoch, we compare the five epochs with the lowest FID scores and select the one that demonstrates a broader distribution of learned parameters. This ensures a balance between a low FID score and diversity in the learned parametric distributions, preventing the selection of a mode-collapsed model and ensuring that the final model produces high-quality and varied synthetic data. 

In Figure~\ref{fig:synth}, we present samples of the synthetic data generated using the trained models for each material case, along with their automatically computed mask images (see Section~\ref{m_inference}). Additional synthetic samples, visualizations of the learned parametric distributions, and an overview of training parameters are provided in the supplementary information.

\begin{table}[h]
    \centering
    \resizebox{1.0\textwidth}{!}{%
    \begin{tabular}{llccc}
    \hline
        \textbf{Domain} & \textbf{Model} & \textbf{DSC} & \textbf{AP\(_{50}\)} & \textbf{PQ} \\
\hline
\multirow{3}{*}{TiO\(_2\) HIM} 
& Model - Real & $\mathbf{0.968 \pm 3.16 \times 10^{-4}}$ & $\mathbf{0.737 \pm 0.014}$ & $\mathbf{0.938 \pm 5.93 \times 10^{-4}}$ \\
& Model - Synth Mill et al. & $0.906 \pm 0.009$ & \(\underline{0.493 \pm 0.020}\) & $0.829 \pm 0.015$ \\
& Model - Synth Ours  &  \(\underline{0.932 \pm 0.003}\) &  $0.393 \pm 0.016$ &  \(\underline{0.874 \pm 0.005}\) \\
\hline
\multirow{3}{*}{SiO\(_2\) HIM} 
& Model - Real & $\mathbf{0.955 \pm 9.49 \times 10^{-4}}$ & $\mathbf{0.945 \pm 0.016}$ & $\mathbf{0.914 \pm 0.002}$ \\
& Model - Synth Mill et al. & $0.786 \pm 0.002$ & $0.375 \pm 0.004$ & $0.659 \pm 0.003$ \\
& Model - Synth Ours & \(\underline{0.860 \pm 4.86 \times 10^{-4}}\) & \(\underline{0.478 \pm 0.011}\) & \(\underline{0.754 \pm 6.21 \times 10^{-4}}\) \\
\hline
\multirow{3}{*}{TiO\(_2\) SEM} 
& Model - Real & $\mathbf{0.964 \pm 0.001}$ & $\mathbf{0.567 \pm 0.012}$ & $\mathbf{0.930 \pm 0.001}$ \\
& Model - Synth R\"uhle et al. & $0.911 \pm 0.001$ & $0.467 \pm 0.017$ & $0.837 \pm 0.002$ \\
& Model - Synth Ours & \(\underline{0.916 \pm 0.003}\) & \(\underline{0.474 \pm 0.033}\) & \(\underline{0.845 \pm 0.004}\) \\
\hline
    \end{tabular}
    } 
    \caption{\textbf{Quantitative Evaluation Results of Segmentation Performance for Real and Synthetic Data.} The table presents the mean and variance of test performance across three runs, measured by the Dice Similarity Coefficient (DSC), Average Precision at 50\% IoU (AP\(_{50}\)), and Panoptic Quality (PQ) for different segmentation models trained on real and synthetic datasets across various domains: TiO\(_2\) in HIM, SiO\(_2\) in HIM, and TiO\(_2\) in SEM. The ``Model - Real" rows represent the averaged test performance of the real-data models. ``Model - Synth Mill et al." refers to models trained on synthetic data generated by Mill et al. (2021) \cite{mill2021synthetic}. Similarly, ``Model - Synth R\"uhle et al." refers to models trained on synthetic data generated by R\"uhle et al. (2021) \cite{ruhle2021workflow}. ``Model - Synth Ours" refers to models trained on synthetic data generated by our DiffRenderGAN approach. Bold values indicate the best scores for each metric within a domain, and underlined values highlight the top scores among synthetic models.}
    \label{tab:evaluation_results}
\end{table}

\section{Deep Learning-Based Segmentation of Nanoparticles Trained on Synthetic Images}\label{Evaluation}
After training DiffRenderGAN on the four image cases, we assessed its effectiveness by training segmentation models on each respective synthetic dataset. For three of these cases (TiO\(_2\) HIM, SiO\(_2\) HIM, and TiO\(_2\) SEM), synthetic data produced by previously published methods is available for comparison \cite{mill2021synthetic}\cite{ruhle2021workflow}. For the AgNW case, where no alternative synthetic data or ground truth annotations are available, we performed a qualitative assessment of our synthetic data to demonstrate its effectivness for rod-like nanoparticles. 

\begin{figure}[t]
\centering
\includegraphics[width=0.7\textwidth]{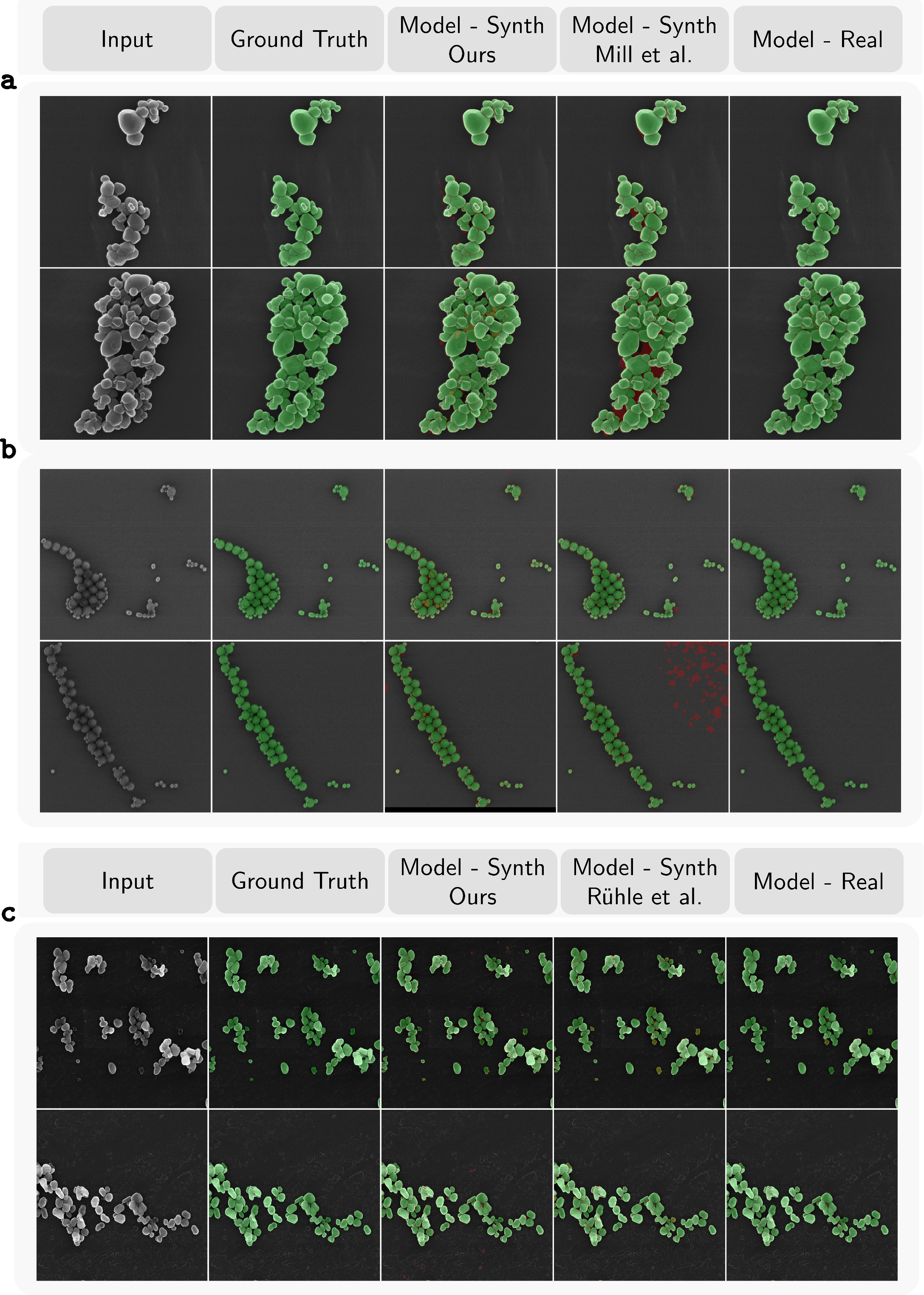}
\caption{\textbf{Excerpt of Segmentation Test Results from Models Trained on Real and Synthetic Data.} 
Input images with overlaying corresponding ground truth masks are compared with the segmentation results from models trained on synthetic data generated by our method, synthetic data from a prior study, and a model trained on real data. 
Green overlay sections represent true positive pixels in comparison to the ground truth mask, red sections indicate false positives (pixels wrongly identified as particles), and yellow highlights show missed particle pixels (false negatives). 
Each image pair is selected based on the run with the highest DSC for each model. \textbf{a.} TiO$_2$ \& \textbf{b.} SiO$_2$ in HIM from Mill et al. (2021) \cite{mill2021synthetic}. \textbf{c.} TiO$_2$ in SEM from Rühle et al. (2021) \cite{ruhle2021workflow}, here cropped for visualization reasons.}
\label{fig:eval}
\end{figure}

To comprehensively evaluate segmentation performance across different aspects, we employ three key metrics: the Dice Similarity Coefficient (DSC), Average Precision (AP) at an Intersection-over-Union (IoU) threshold of 50\%, and Panoptic Quality (PQ). The DSC measures the overlap between predicted and ground truth segmentation masks \cite{muller2022towards}, providing a direct assessment of segmentation accuracy. AP quantifies the precision of object localization at a fixed IoU threshold, reflecting a model's ability to correctly detect and delineate nanoparticles \cite{lin2014microsoft}. Lastly, PQ integrates both segmentation and object detection accuracy into a single metric, offering an evaluation of both detection and segmentation performance \cite{kirillov2019panoptic}.
During testing on the remaining 20\% split of the data, we intentionally limited postprocessing to binarization and connected-components analysis to ensure an accurate quality assessment of the synthetic datasets. Our primary objective here was to evaluate the raw segmentation capabilities of models trained on these datasets. Postprocessing techniques can compensate for quality gaps in the synthetic data. For example, watershed-based postprocessing can mitigate the issue of overlapping particles that remain connected during testing. Additionally, we benchmark the synthetic data models, except in the AgNW case, against the test performance of a model trained on real data, which serves as a desired performance reference for the synthetic data models. Quantitative results for the three comparable cases are presented in Table \ref{tab:evaluation_results}, while visual segmentation results are shown in Figure \ref{fig:eval}. The qualitative visual results for AgNW in SEM are provided separately in Figure \ref{fig:eval_ag}. For details on the evaluation procedure, refer to \ref{m_eval}.

\subsection{TiO\(_2\) HIM}
In the TiO\(_2\) HIM case, the segmentation model trained on real data achieved the best results, closely matching the ground truth with a DSC of $0.968 \pm 3.16 \times 10^{-4}$, AP\(_{50}\) of $0.737 \pm 0.014$, and PQ of $0.938 \pm 5.93 \times 10^{-4}$. Among the models trained on synthetic data, our model outperforms the one trained on synthetic data from Mill et al., achieving a DSC of $0.932 \pm 0.003$ compared to $0.906 \pm 0.009$ and a PQ of $0.874 \pm 0.005$ versus $0.829 \pm 0.015$, indicating better segmentation accuracy. However, Mill et al.'s model achieved a higher AP\(_{50}\) score ($0.493 \pm 0.020$ vs. $0.393 \pm 0.016$), suggesting better precision in particle localization and separation. Figure~\ref{fig:eval}a shows that the model trained on real data most closely matches the ground truth images. The model trained on Mill et al.'s synthetic data tends to oversegment, introducing frequent false positives, but excels at distinguishing individual instances of nanoparticles. In contrast, our model trained on synthetic data displays false negatives, such as partially unfilled particles, but introduces fewer false positives. However, it struggles with the separation of nanoparticles, particularly with smaller instances, as indicated by the quantitative results. 

\subsection{SiO\(_2\) HIM}
In the SiO\(_2\) HIM case, the model trained on real data demonstrated superior performance once again, achieving a DSC of $0.955 \pm 9.49 \times 10^{-4}$, AP\(_{50}\) of $0.945 \pm 0.016$, and a PQ of $0.914 \pm 0.002$. Among the models trained on synthetic data, our synthetic data model performed better than Mill et al.'s approach across all metrics. Our model achieved a higher DSC ($0.860 \pm 4.86 \times 10^{-4}$ vs. $0.786 \pm 0.002$), AP\(_{50}\) ($0.478 \pm 0.011$ vs. $0.375 \pm 0.004$), and PQ ($0.754 \pm 6.21 \times 10^{-4}$ vs. $0.659 \pm 0.003$), indicating a better representation of real-world SiO\(_2\) HIM data. Qualitatively (see Figure~\ref{fig:eval}b), the results from the real data models correspond most closely to the ground truth images. In contrast, our synthetic data model effectively identifies true positives but introduces false positives, particularly in the form of small particle instances. Mill et al.'s synthetic data model struggles with both true positive identification and the avoidance of frequent false positives.

\subsection{TiO\(_2\) SEM}
In the TiO\(_2\) SEM domain, among all models, the segmentation model trained on real data achieved the highest performance, with a DSC of \(0.964 \pm 0.001\), a PQ of \(0.930 \pm 0.001\), and an AP\(_{50}\) score of \(0.567 \pm 0.012\). The models trained on synthetic data performed very similarly, with our model achieving a DSC of \(0.916 \pm 0.003\) and a PQ of \(0.845 \pm 0.004\), while Rühle et al.'s model reached a DSC of \(0.911 \pm 0.001\) and a PQ of \(0.837 \pm 0.002\). Their AP\(_{50}\) score were also closely matched, at \(0.474 \pm 0.033\) for our model and \(0.467 \pm 0.017\) for Rühle et al.'s. These results indicate that both synthetic models offer comparable segmentation accuracy and instance detection, with only marginal variations. The qualitative results in Figure~\ref{fig:eval}c further illustrate this similarity, showing that both synthetic models effectively segment particles and handle particle separations in a comparable manner. While slight differences in segmentation behavior exist, such as our model’s tendency for oversegmentation, both approaches perform nearly equivalently. 

\subsection{AgNW SEM}
\begin{figure}[t]
\centering
\includegraphics[width=1.\textwidth]{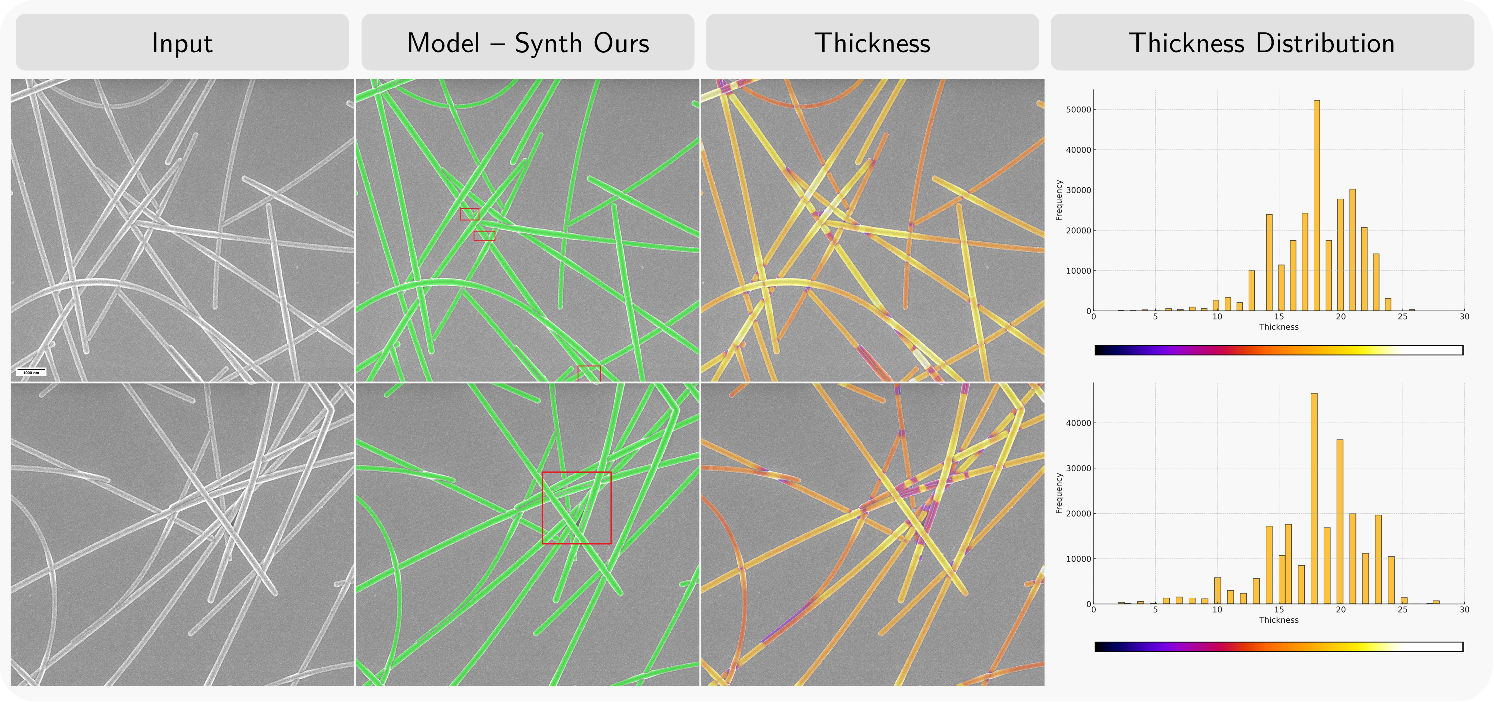}
\caption{\textbf{Qualitative Evaluation of Segmentation Performance and Local Thickness Estimation for AgNW in SEM.} This figure presents a visual analysis of AgNW segmentation and local thickness estimation using a segmentation model trained exclusively on synthetic data generated by the DiffRenderGAN framework. The left column displays two randomly selected SEM images of AgNWs used for testing. The second column presents a green overlay with segmentation results, with notable false segmentations highlighted in red. The third column shows a local thickness estimation overlaid as a heatmap based on the segmentation results, where brighter values represent thicker regions. The right column depicts the corresponding local thickness distributions for each scan, providing insights into morphological variations and potential applications in nanotechnology.}
\label{fig:eval_ag}
\end{figure}

For the AgNW in SEM case, we conducted a qualitative evaluation due to the absence of annotated ground truth data. Figure~\ref{fig:eval_ag} presents the segmentation performance of a model trained on synthetic AgNW images generated by DiffRenderGAN. Overall, the model performs well in identifying nanowires, though frequent false negatives can be oberserved. Despite these errors, the model effectively segments nanowires, highlighting the potential of our synthetic data for this application. Figure~\ref{fig:eval_ag} also demonstrates an example application of integrating DiffRenderGAN’s framework into nanowire image quantification. Specifically, we apply local thickness calculations to examine the rod-like structure of the AgNWs. The figure shows an overlay of local thickness measurements based on the model’s segmentations, along with the corresponding thickness distributions.

\section{Conclusion}
In this study, we present DiffRenderGAN, a novel generative model designed for generating synthetic training data for microscopy analysis. By integrating differentiable rendering into a GAN, DiffRenderGAN directly addresses the challenges posed by the limited availability of (annotated) data, which represents a significant bottleneck in training deep learning models for segmentation and analysis in microscopy research.

We assessed DiffRenderGAN across various material morphologies and modalities, including TiO\(_2\) and SiO\(_2\) in HIM, TiO\(_2\) in SEM, and AgNW for rod-like particles in SEM. Our results showed that while models trained on real data consistently achieved the highest performance across segmentation metrics, DiffRenderGAN’s synthetic data often matched or exceeded the performance of existing synthetic data techniques. Despite some challenges, particularly in improving particle segregation, DiffRenderGAN shows promise in generating high-quality synthetic data for segmentation-based applications across multimodal microscopy domains, effectively narrowing the domain gap between synthetic and real data. Additionally, it offers a simplified and streamlined approach using a single model for image generation, avoiding multi-stage model training or time-intensive expert-guided rendering methods. This limits manual intervention to providing basic nanoparticle meshes and selecting a scale and positional strategy for realistic nanoparticle mesh alignment.

Looking ahead, DiffRenderGAN’s potential applications extend beyond HIM and SEM. Future research should explore its use with other imaging techniques, such as Atomic Force Microscopy (AFM) and Computed Tomography (CT), as well as across a broader range of nanomaterials. Such studies will test the generalizability of DiffRenderGAN and could unlock its potential to accelerate research in fields where the availability of representative data remains a critical bottleneck.

In conclusion, DiffRenderGAN represents a substantial advancement in synthetic data generation for microscopy, offering an efficient, scalable, and integrated solution. Although a representativeness gap compared to real data remains, DiffRenderGAN significantly reduces this gap, paving the way for more robust and comprehensive image-based analyses in the study of complex nanomaterial systems.

\section{Methods}\label{sec11}
\subsection{Image Acquisition and Processing}\label{method_data}
Details regarding sample preparation can be found in the respective publications \cite{mill2021synthetic}\cite{goebelt2015encapsulation}\cite{ruhle2021workflow}. 

\subsubsection{AgNW SEM} Data for silver nanowires were provided by the authors of ref. \cite{goebelt2015encapsulation} (personal communication, unpublished). The sample consisted of AgNWs synthesized according to Korte et al. (2008) \cite{korte2008rapid}, which were drop-cast onto a silicon substrate and coated with 100 nm of aluminum-doped zinc oxide (AZO) via atomic layer deposition (ALD), indium-free electrode. Details regarding the sample preparation and the use of this material can be found in the respective publication by Göbelt et al. (2015) \cite{goebelt2015encapsulation}.
The sample was imaged using a Zeiss MultiSEM 505 multibeam SEM at a landing energy of 3 keV. The MultiSEM 505 employs 61 primary electron beams for parallel SEM imaging of large sample surfaces at high resolution \cite{eberle2015high}. For this sample, the step size was set to 10 nm and the images were 1252 $\times$ 1092 pixels. In total, 10431 individual images (171 locations $\times$ 61 beams) with a combined size of more than 14 GPixel and an area of approx. 1 mm² were scanned within less than 5 minutes. From the dataset, twelve images were randomly selected for this study. Of these, ten images were used for training DiffRenderGAN, while the remaining two images were reserved for testing.

\subsubsection{TiO\(_2\) SEM}
Rühle et al. 2021 used TiO2 nanomaterial from the Horizon 2020 project \href{www.acena no-proje ct.eu/}{ACEnano}, which was ultrasonicated in ultrapure water and drop cast onto conventional carbon TEM grids and subsequently analyzed using a FEG-SEM (Supra 40, Zeiss) with an In-Lens detector and in transmission mode (not used in DIffRenderGAN evaluation). Real data was sourced from the repository specified in the supplementary section of Rühle et al. (2023), comprising 40 SEM images with 1024 $\times$ 768 pixels. Out of these, 32 images were used for training our and Rühle et al's approach. While eight images were reserved for testing, 1,000 annotated synthetic images were generated using the available software from their repository \cite{ruhle2021workflow}. Each synthetic image had a resolution of 512 × 312 pixels and was created following the original procedure.

\subsubsection{TiO\(_2\) and SiO\(_2\) HIM}
In Mill et al., SiO2 nanoparticles with two different diameters and food grade TiO2 nanoparticles (E171) with a size distribution of 20 to 240 nm, both deposited on silicon chips (reference AGAR: G3390-10), were obtained from the “Laboratoire National de métrologie et d’Essais”. Secondary electron images of the particles were obtained on a Zeiss ORION NanoFab using the helium ion beam at an energy of 25 keV and a beam current of 0.5 pA. The NanoFab used a side-mounted secondary electron detector similar to an Everhardt-Thornley type detector.  Synthetic datasets included 180 SiO\(_2\) and 180 TiO\(_2\) annotation-paired images at a resolution of 2031 $\times$ 2031 pixels. Additionally, eight real TiO\(_2\) (six used for training DiffRenderGAN, two reserved for testing) and nine real SiO\(_2\) (seven used for training DiffRenderGAN, two reserved for testing) annotation-paired images, each with a resolution of 2031 $\times$ 2031 pixels, were provided upon request. 

\subsection{DiffRenderGAN Framework}

\subsubsection{Model Design}\label{m_model_design}
Our GAN model employs a three-layer PatchGAN discriminator based on the CycleGAN architecture \cite{CycleGAN2017}. The generator consists of three key modules:

\begin{enumerate}
    \item \textbf{Regression Model}: A five-layer deep neural network, where the first four layers consist of \textit{Dropout}, a \textit{Fully-Connected Layer} (\textit{in} = 128, \textit{out} = 128), and \textit{ReLU} activation. The final layer is a \textit{Fully-Connected Layer} with \textit{Sigmoid} activation, responsible for regressing BSDF parameters and the noise scale. \textit{Weight normalization} is applied across all layers.
    
    \item \textbf{Differentiable Rendering Function}: Utilizing Mitsuba 3.4 \cite{Mitsuba3}, this module processes the current virtual scene state and the parameters predicted by the regression model to generate rendered images.
    
    \item \textbf{Noise-Adding Function}: After rendering, zero-centered scalable Gaussian noise is added to the images to simulate realistic imaging conditions.
\end{enumerate}

\subsubsection{Virtual Scene Design}\label{m_virtual_scene}
Before conducting experiments, a virtual scene (utilized by Mitsuba 3 \cite{Mitsuba3}) was designed. This scene consists of a rectangular mesh acting as a stage, a toroidal light source surrounding the stage mesh, and a camera aligned perpendicularly to the center of the stage. An additional rectangular mesh light source is positioned above the camera. Nanoparticle meshes should be centrally located within the torus and stage mesh and are dynamically scaled and translated during training. After each image generation, the nanoparticle meshes reset to their initial position in the center of the stage mesh. The scene is rendered using a \textit{Perspective Sensor}, a \textit{Stratified Sampler}, and a \textit{Gaussian Reconstruction Filter}. Both the stage and the nanoparticle meshes utilize a \href{https://mitsuba.readthedocs.io/en/stable/src/generated/plugins_bsdfs.html#the-principled-bsdf-principled}{\textit{Principled BSDF}}. All BSDF parameters not involved in the optimization process remain fixed at their default values. The scene integrator is set to \textit{Direct Reparam}, and the \textit{area light} plugin is used for both emitters. The optimized BSDF parameters for the nanoparticle mesh in the experiments include \textit{Roughness}, \textit{Base Color}, \textit{Sheen}, \textit{Sheen Tint}, and \textit{Specular Tint}, while the stage BSDF optimizes only the \textit{Base Color} parameter. Emission values remain fixed during optimization, with the toroid mesh intensity set to 1.0 and the rectangle mesh intensity set to 0.1.

\subsubsection{Nanoparticle Mesh Modeling}\label{m_mesh}
Prior to training DiffRenderGAN, nanomaterial meshes were modeled using Blender version 3.6. Predesigned meshes were adapted to match the morphologies of real particles across different domains: cubes for TiO\(_2\), cones for Ag, and spheres for SiO\(_2\). For each material, a collection of meshes was duplicated and, where necessary, deformed through bending, vertex translation, and rotation. 

The following configurations were used in our experiments:
\begin{itemize}
    \item 10 bent and randomly rotated cones for AgNW
    \item 20 non-deformed spheres for SiO\(_2\) HIM
    \item 15 deformed and smoothed cubes for TiO\(_2\) HIM
    \item 40 deformed and smoothed cubes for TiO\(_2\) SEM
\end{itemize}

To achieve smoother mesh surfaces, the Blender \textit{Remesh} modifier was applied. Each mesh was positioned at the origin (0, 0, 0) and exported using the \href{https://github.com/mitsuba-renderer/mitsuba-blender}{Mitsuba-Blender} plugin to ensure compatibility with the differentiable rendering software. This workflow was designed to require minimal expertise in 3D rendering.

\subsubsection{Transformation Computation}\label{m_trafo}
To simulate realistic particle sizes and spatial distributions in synthetic images, a size distribution (bimodal, lognormal, or normal) and a spatial arrangement model (random or agglomerated) are selected based on expert analysis of real images prior to training. Once the size and positional distributions are chosen, along with the synthetic image sample size, a transformation tensor is computed. This tensor is generated by sampling from the selected size and positional distributions for each image optimized during training, assigning a scaling factor and positional coordinates to each mesh.

It is critical that the size and spatial distribution model parameters align with the exported sizes of the nanoparticle meshes. Otherwise, synthetic images may contain nanoparticles that are either too large or too small. For random spatial arrangements, the meshes are uniformly distributed within the virtual scene, either in a planar configuration (e.g., AgNW SEM) or in three-dimensional space. For agglomerated arrangements, a Poisson disk-based sampling algorithm was employed to simulate clusters \cite{bridson2007fast}.

\subsubsection{Model Training}\label{m_trian}
All DiffRenderGAN models were trained using PyTorch \cite{paszke2019pytorch}. The generator's learning rate was set to 0.0002, while the discriminator's learning rate was set to 0.0001, both optimized using Adam \cite{kingma2017adammethodstochasticoptimization}. Xavier initialization was applied to both the generator and discriminator \cite{glorot2010understanding}. For all experiments, DiffRenderGAN was trained on $256 \times 256$ pixel image patches for 50 epochs. Each training dataset was cropped into overlapping patches of size $256 \times 256$ pixels. DiffRenderGAN was then trained on image patches containing at least three fully displayed particles while avoiding repetitive particle patches. Each experiment utilized a batch size of 1. The image patches included in the training were as follows: 82 for AgNW, 56 for SiO\(_2\) in HIM, 126 for TiO\(_2\) in HIM, and 124 for TiO\(_2\) in SEM. To monitor the quality of synthetic images, after each epoch, a synthetic dataset matching the size of the real dataset was generated, and the Fréchet Inception Distance (FID) score was calculated. The best epoch was selected as detailed in the main text.

\subsubsection{Model Inference}\label{m_inference}
After training, the generator was loaded with the respective best epoch state. During runtime, an additional duplicate scene, without the stage mesh, was created, where an \textit{AOV} integrator however was used. This integrator enables the rendering of labeled images displaying unique identifiers for each mesh observed in the camera's field of view, necessitating the removal of the stage mesh. The generated label images were processed through rounding to ensure discrete label values. Subsequently, these label images were binarized, and an additional contour class was added. For TiO\(_2\) and SiO\(_2\) in HIM and AgNW in SEM, a contour thickness of four pixels was used, while for TiO\(_2\) in SEM, a thickness of one pixel was applied. To ensure meaningful synthetic images during inference, only images where particles exhibited sufficient contrast against the background were rendered. Specifically, based on the mask information, we automatically removed images during inference where the mean intensity of the particles was less than 15\% in comparison to the mean intensity of the background. Following this automated strategy during inference, each experiment produced 1,000 paired synthetic images with their respective annotated masks.

\subsection{Workflow for Deep Learning-Based Segmentation of Nanoparticles}\label{m_eval}
For performance comparisons, we employed the nnUNet framework \cite{isensee2021nnu}, which automatically configures model parameters based on the characteristics of each individual dataset. This approach eliminates potential performance bias that could arise from manual model selection and configuration. All segmentation models were trained for multiclass segmentation using nnUNet's default training procedure, classifying pixels into three categories: particle, contour, and background. The contour class specifically aids in distinguishing overlapping particles. During the addition of the contour class for real data and synthetic data from other methods, we ensured that no particle information in the respective masks was overwritten by the addition of contour class pixels.

To ensure robust evaluation of our models and mitigate potential biases introduced by artifact-rich environments in original scans, we supplemented our synthetic image datasets with 200 real background patches (i.e., images without particles) randomly sampled from overlapping patches of the corresponding real training data, used during DiffRenderGAN training. The rationale behind this supplementation was to encourage our synthetic data models to accurately distinguish true particles from irrelevant artifacts, such as dirt or preparational anomalies, during segmentation tasks. Since our proposed method does not provide additional meshes for artifacts and only generates "clean" images, this strategy introduces additional robustness under varying imaging conditions. We note that this aspect is explicitly or implicitly considered across all approaches: Mill et al. supplemented their data by including dirt textures as synthetic backgrounds, partially addressing artifact-related challenges. Rühle et al.'s GAN-based approach inherently integrates the generation of background and artifacts as long as they are present in the training dataset. Therefore, we follow these methods as proposed by the original authors.

The datasets used for nnUNet training included the following:

\begin{itemize}
    \item \textbf{TiO\(_2\) HIM}: Our segmentation model was trained on our synthetic dataset (1,000 images, $256 \times 256$ pixels) supplemented with real background patches (200 images, $256 \times 256$ pixels; total: 1,200 images). Mill et al.'s model was trained on their synthetic dataset (180 images, $2,031 \times 2,031$ pixels). The real-data model was trained on 294, $256 \times 256$ image patches extracted from six real images in the training split, utilizing the available respective ground truth masks for training.

    \item \textbf{SiO\(_2\) HIM}: Our segmentation model was trained on our synthetic dataset (1,000 images, $256 \times 256$ pixels) supplemented with real background patches (200 images, $256 \times 256$ pixels; total: 1,200 images). Mill et al.'s model was trained on their synthetic dataset (180 images, $2,031 \times 2,031$ pixels). The real-data model was trained on 343, $256 \times 256$ image patches extracted from seven real images in the training split, utilizing the available respective ground truth masks for training.

    \item \textbf{TiO\(_2\) SEM}: Our segmentation model was trained on our synthetic dataset (1,000 images, $256 \times 256$ pixels) supplemented with real background patches (200 images, $256 \times 256$ pixels; total: 1,200 images) and tested on eight real images. Rühle et al.'s model, trained on their synthetic dataset (1,000 images, $512 \times 312$ pixels). The real-data model was trained on 256, $256 \times 256$ image patches extracted from 32 real images in the training split, utilizing the available respective ground truth masks for training.

    \item \textbf{AgNW SEM}: Our segmentation model was trained on our synthetic dataset (1,000 images, $256 \times 256$ pixels) supplemented with real background patches (200 images, $256 \times 256$ pixels; total: 1,200 images).
\end{itemize}

Each model was trained for three runs. After training, each model from each run within its respective domain was tested on the same real test data (TiO\(_2\) HIM: two image scans; SiO\(_2\) HIM: two image scans; TiO\(_2\) SEM: eight image scans). After testing, each image was binarized using only the particle class information. We then computed the mean test performance for each model within each run and calculated the mean and variance of the test performance across all runs using the evaluation metrics introduced in Section~\ref{Evaluation}.

\subsection{Technical Notes}\label{method_tech}
All experiments involving DiffRenderGAN and nnUNet were conducted using Python 3.9.18 on an NVIDIA A40 GPU with CUDA 12.3.

\backmatter


\bibliography{sn-bibliography}

\section*{Acknowledgments}
S.C. was supported by the European Union’s H2020 research and innovation program under the Marie Sklodowska-Curie grant agreement AIMed ID: 861138. D.P, D.A., G.S. and S.C. acknowledge the financial support from the European Union within the research projects 4D + nanoSCOPE ID: 810316, LRI ID: C10, STOP ID: 101057961, from the German Research Foundation (DFG) within the research project UNPLOK ID: 523847126, and from the “Freistaat Bayern” and European Union within the project Analytiktechnikum für Gesundheits- und Umweltforschung AGEUM, StMWi-43-6623-22/1/3.

\newpage
\pagestyle{empty}
\section*{Supplementary Information}
\setcounter{figure}{0}   
\renewcommand{\figurename}{Fig. S}
\setcounter{table}{0}   
\renewcommand{\tablename}{Table S}

\begin{sidewaystable}
    \centering
    \begin{tabular}{l|ccccccc}
        \hline
        \textbf{Material}  
        & \textbf{Train Images} 
        & \textbf{Best Epoch} 
        & \textbf{FID Score} 
        & \textbf{Noise Limits} 
        & \textbf{Stage BSDF Limits} 
        & \textbf{Particle BSDF Limits}
        & \textbf{Duration} \\
        \hline
        TiO$_2$ (HIM) & 126 & 14 & 126 & 0.001--0.2 & 0.01--0.3 & 0.01--5.0 & 1h28m \\
        TiO$_2$ (SEM) & 124 & 28 & 134 & 0.001--0.1 & 0.01--0.4 & 0.01--5.0 & 1h28m \\
        AgNW (SEM)    & 82  & 43 & 195 & 0.001--0.2 & 0.01--5.0 & 0.01--5.0 & 31m   \\
        SiO$_2$ (HIM) & 56  & 49 & 144 & 0.001--0.2 & 0.01--0.3 & 0.01--5.0 & 41m3 \\
        \hline
    \end{tabular}
    \caption{\textbf{DiffRenderGAN Experimental Overview.} This table summarizes the DiffRenderGAN experiments for each image/material case. 
    The “best epoch” was determined by selecting, among the five epochs with the lowest FID scores, the one in which the model exhibited the broadest parameter distribution (see e.g. Figure S\ref{s_tio22_dist}). 
    This chosen best epoch was then used for further processing and for generating the synthetic datasets used to investigate the segmentation capabilities of models trained on our synthetic data (see main text). For each model, the table lists the FID score at the best epoch, the size of the training dataset, the noise, stage \& particle mesh BSDF limits used to rescale the sigmoid outputs \([0,1]\) of DiffRenderGAN's parameter estimation model to the BSDF and noise target domains, and the total training duration.}\label{stable_overview}
\end{sidewaystable}

\begin{figure}[H]
\centering
\includegraphics[width=1.\textwidth]{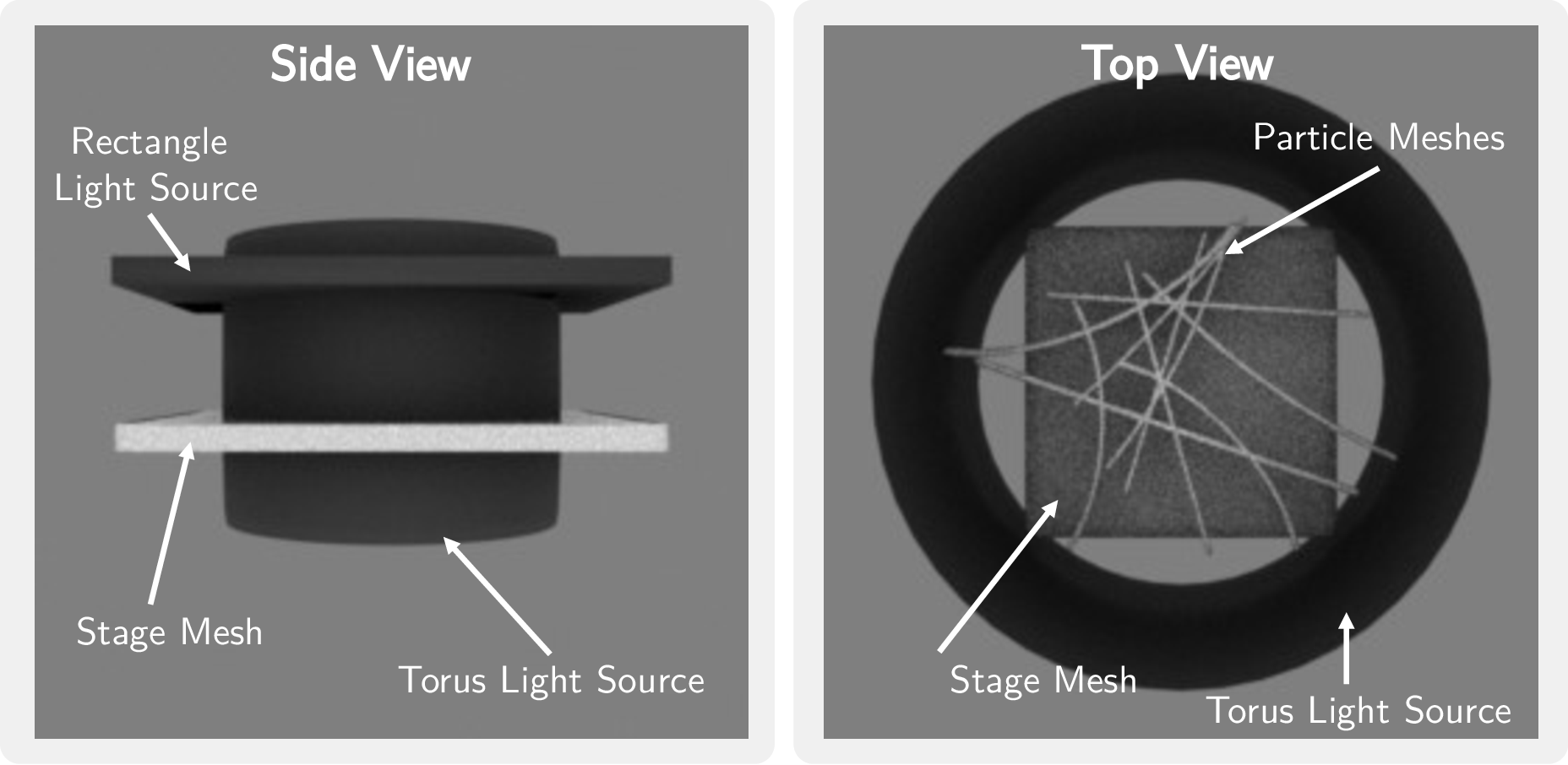}
\caption{\textbf{Virtual Scene Structure used in DiffRenderGAN Experiments.} Here, for visualization, emissions are omitted, and the stage and rectangle light source meshes are scaled in the side view. The stage and nanoparticle meshes are enclosed within a toroidal mesh that acts as a light source, creating glowing edge effects similar to those seen in ion or electron microscopy.}
\end{figure}

\begin{figure}[H]
\centering
\includegraphics[width=1.\textwidth]{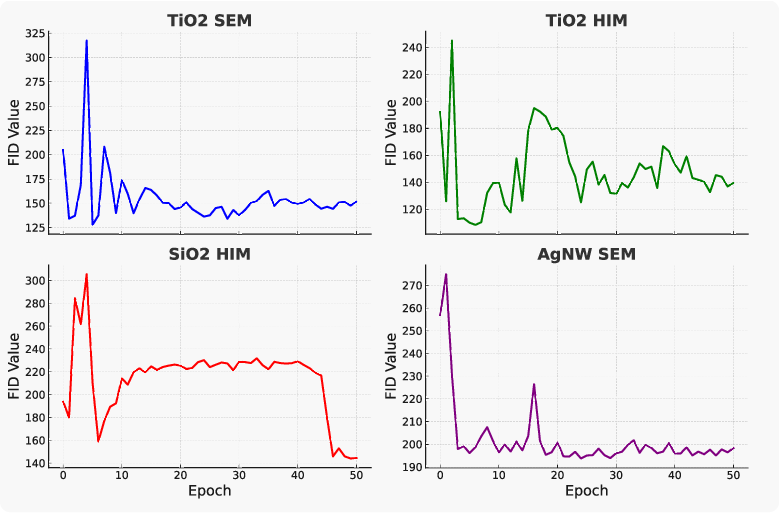}
\caption{\textbf{DiffRenderGAN Training FID Score Progression Over 50 Training Epochs.} This figure shows the FID scores for the four experiments discussed in the main text.
Lower FID values indicate greater dataset similarity between generated and real images.}
\end{figure}

\begin{figure}[t]
\centering
\includegraphics[width=1.\textwidth]{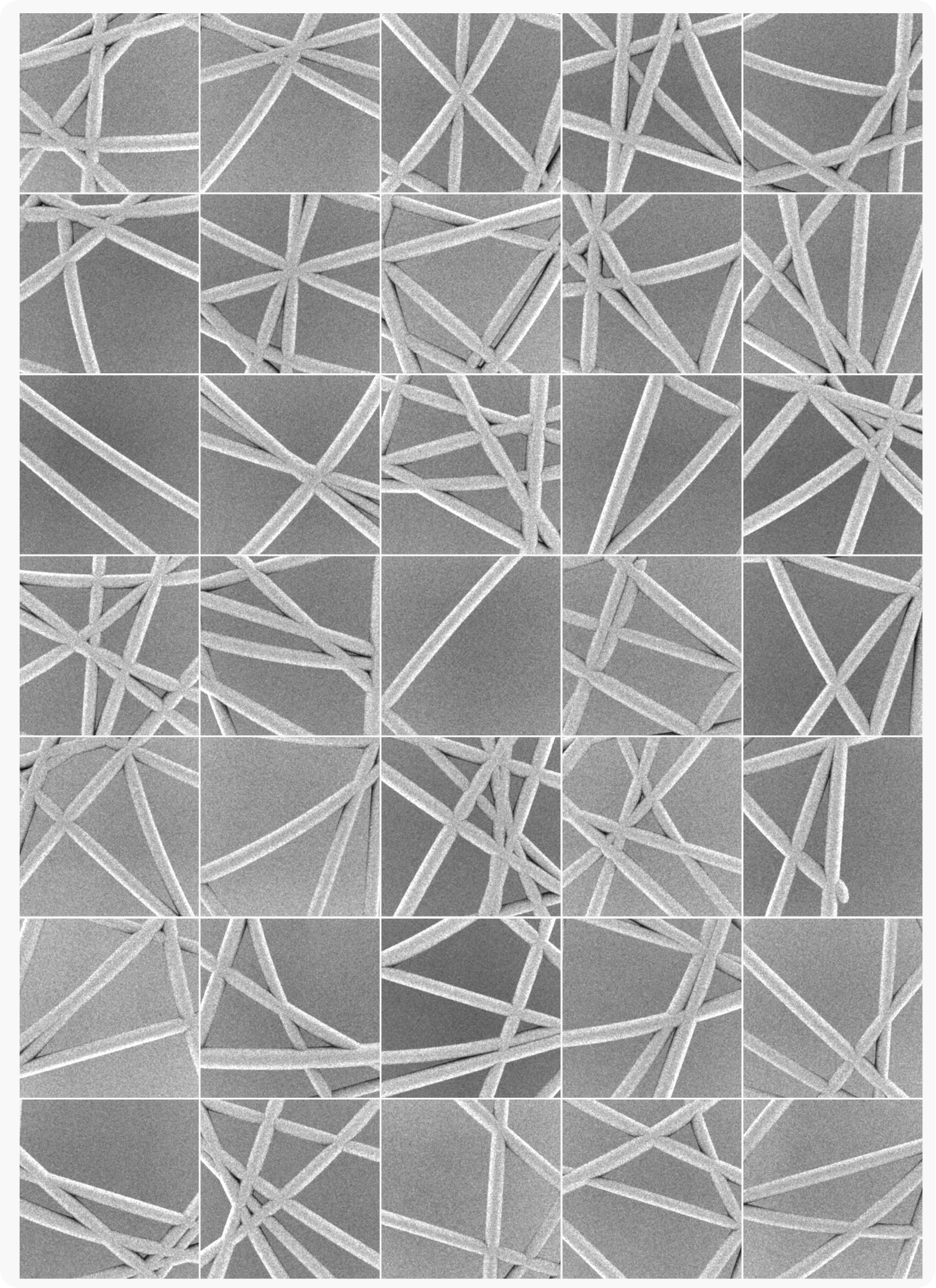}
\caption{\textbf{AgNW in SEM - Example Synthetic Images generated by DiffRenderGAN after Training.} These synthetic images were produced using the model selected at the best epoch (see Table~S\ref{stable_overview}), with 10 cone meshes used in the virtual scene.}
\end{figure}

\begin{figure}[t]
\centering
\includegraphics[width=1.\textwidth]{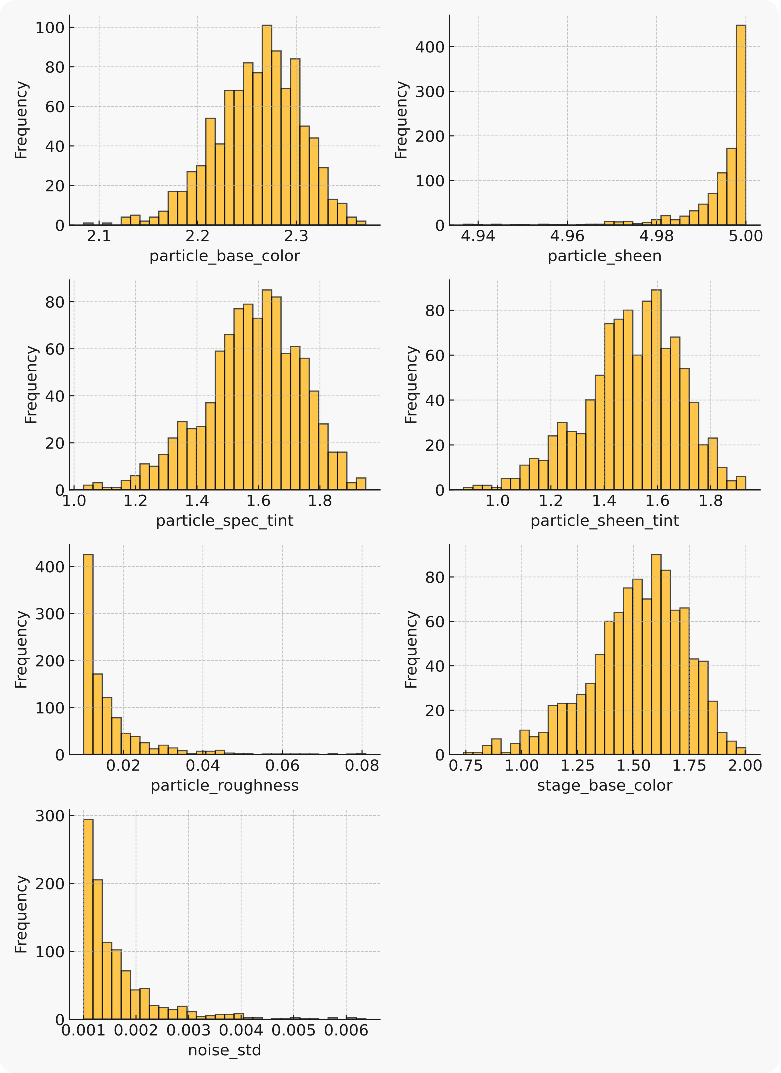}
\caption{\textbf{AgNW in SEM - Histograms of Optimized Scene Parameters after DiffRenderGAN Training.} The histograms in the figure show the distributions of the stage and particle mesh BSDF parameters, as well as the noise deviation, obtained by sampling 1,000 times from the best epoch model state.}
\end{figure}

\begin{figure}[t]
\centering
\includegraphics[width=1.\textwidth]{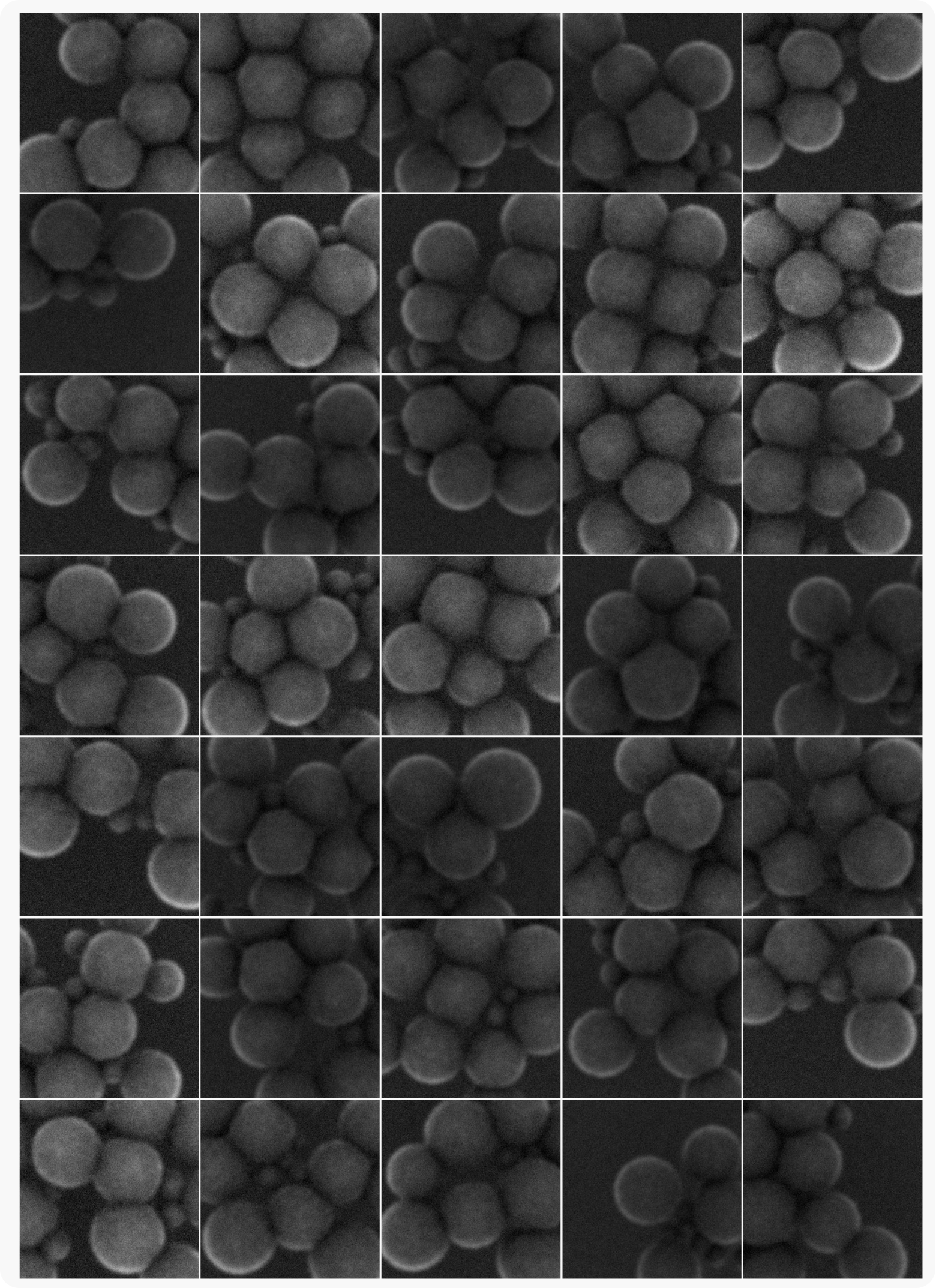}
\caption{\textbf{SiO\(_2\) in HIM - Example Synthetic Images generated by DiffRenderGAN after Training.} These synthetic images were produced using the model selected at the best epoch (see Table~S\ref{stable_overview}), with 20 sphere meshes used in the virtual scene.}
\end{figure}

\begin{figure}[t]
\centering
\includegraphics[width=1.\textwidth]{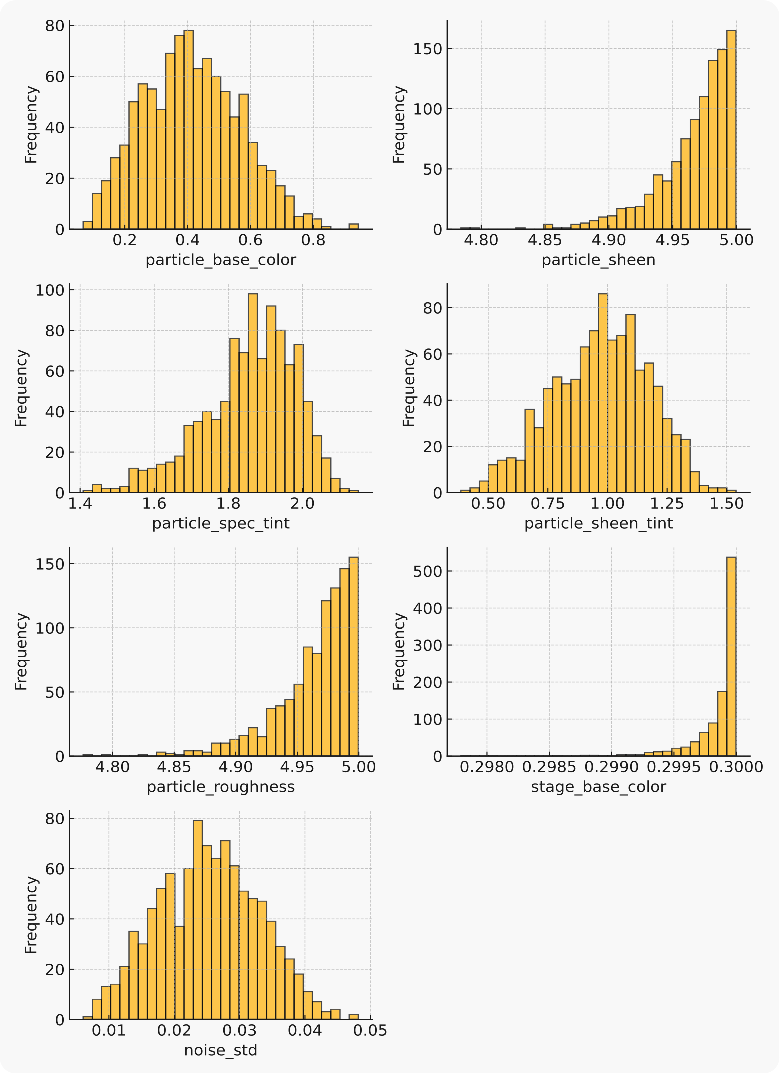}
\caption{\textbf{SiO\(_2\) in HIM - Histograms of Optimized Scene Parameters after DiffRenderGAN Training.} The histograms in the figure show the distributions of the stage and particle mesh BSDF parameters, as well as the noise deviation, obtained by sampling 1,000 times from the best epoch model state.}
\end{figure}

\begin{figure}[t]
\centering
\includegraphics[width=1.\textwidth]{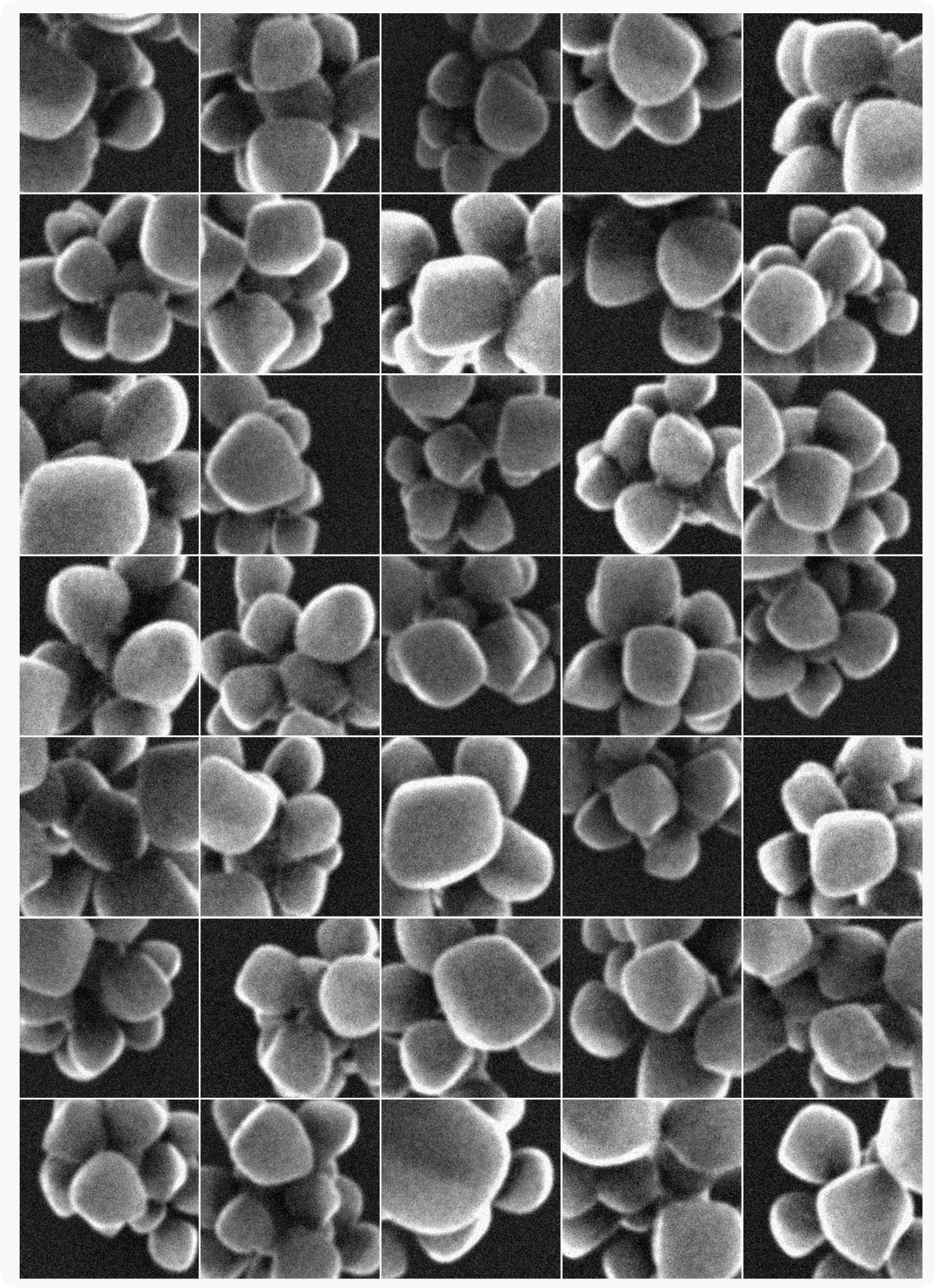}
\caption{\textbf{TiO\(_2\) in HIM - Example Synthetic Images generated by DiffRenderGAN after Training.} These synthetic images were produced using the model selected at the best epoch (see Table~S\ref{stable_overview}), with 15 cube-based meshes used in the virtual scene.}
\end{figure}

\begin{figure}[t]
\centering
\includegraphics[width=1.\textwidth]{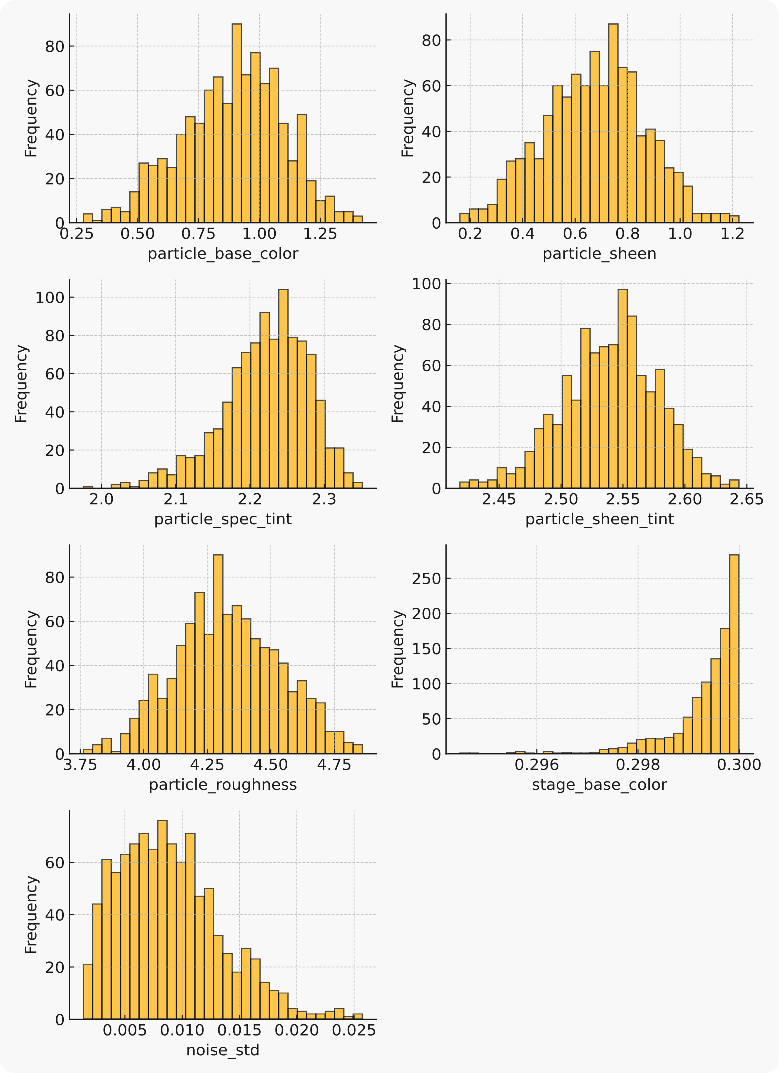}
\caption{\textbf{TiO\(_2\) in HIM - Histograms of Optimized Scene Parameters after DiffRenderGAN Training.} The histograms in the figure show the distributions of the stage and particle mesh BSDF parameters, as well as the noise deviation, obtained by sampling 1,000 times from the best epoch model state.}
\end{figure}

\begin{figure}[t]
\centering
\includegraphics[width=1.\textwidth]{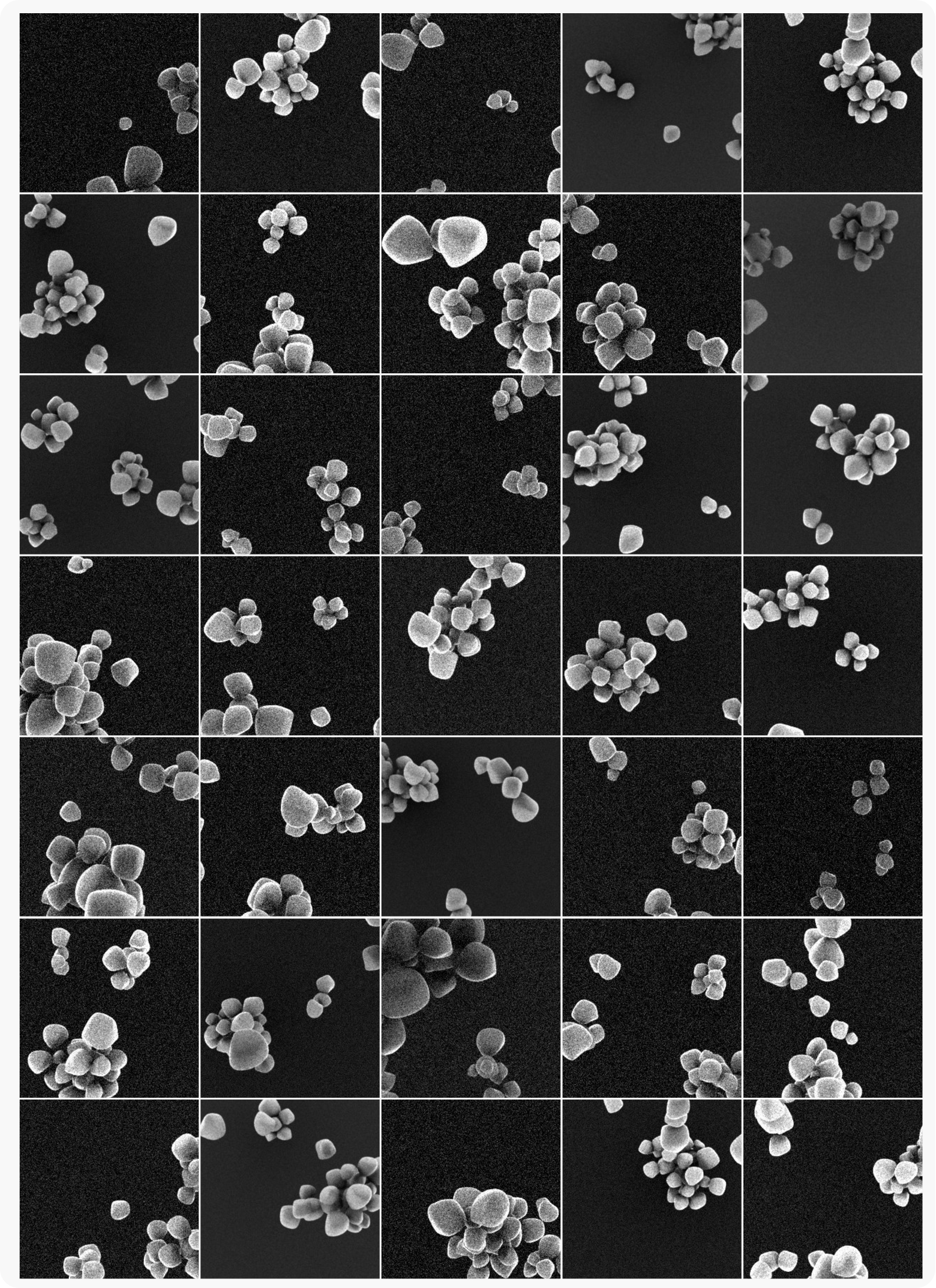}
\caption{\textbf{TiO\(_2\) in SEM - Example Synthetic Images generated by DiffRenderGAN after Training.}These synthetic images were produced using the model selected at the best epoch (see Table~S\ref{stable_overview}), with 40 cube-based meshes used in the virtual scene.}
\end{figure}

\begin{figure}[t]
\centering
\includegraphics[width=1.\textwidth]{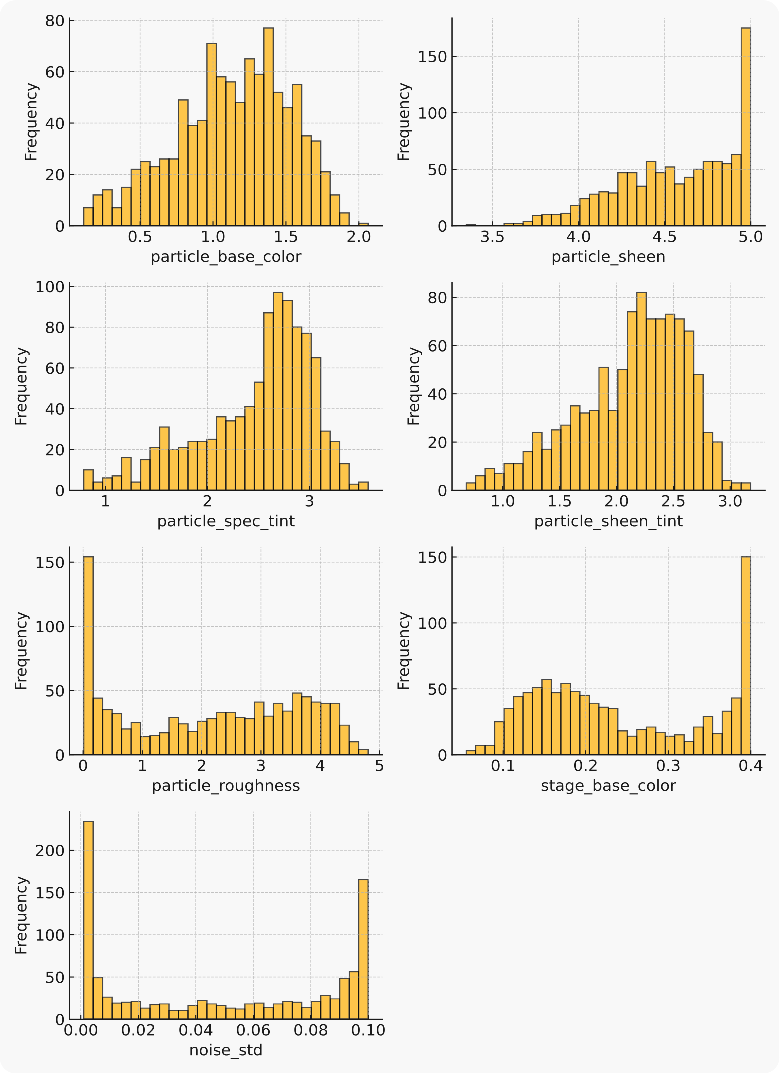}
\caption{\textbf{TiO\(_2\) in SEM - Histograms of Optimized Scene Parameters after DiffRenderGAN Training.} The histograms in the figure show the distributions of the stage and particle mesh BSDF parameters, as well as the noise deviation, obtained by sampling 1,000 times from the best epoch model state.}
\label{s_tio22_dist}
\end{figure}

\end{document}